\documentstyle[11pt]{article}

\def\kon#1#2{\vbox{\halign{##&&##\cr\lower4pt
\hbox{$\scriptscriptstyle\vert$}\hrulefill &\hrulefill\lower4pt
\hbox{$\scriptscriptstyle\vert$}\cr $#1$&$#2$\cr}}}

\def\al{\alpha}

\def\ro{\varrho}

\def\eh{{\scriptstyle{1\over 2}}}
\def\d{\partial}
\def\=d{\,{\buildrel\rm def\over =}\,}

\def\sqr#1#2{{\vcenter{\vbox{\hrule height.#2pt\hbox{\vrule width.
#2pt height#1pt \kern#1pt \vrule width.#2pt}\hrule height.#2pt}}}}

\def\te{\vartheta}
\def\B{\Bigl}

\begin{document}

\title{Nonstandard general relativity III: The Kerr class}
\author{G\"unter Scharf \footnote{e-mail: scharf@physik.uzh.ch}
\\ Institut f\"ur Theoretische Physik, 
\\ Universit\"at Z\"urich, 
\\ Winterthurerstr. 190 , CH-8057 Z\"urich, Switzerland}

\date{}

\maketitle\vskip 3cm

\begin{abstract}  
We revisit the Kerr metric in Boyer-Lindquist coordinates and construct the corresponding class of nonstandard solutions of Einstein's equations. These solutions can be used to describe the outer part of spiral galaxies without assuming dark matter.

\end{abstract}

\newpage

\section{Introduction}

Standard general relativity is the marriage between geometry and gravity. Indeed, geometry is such a pretty woman that it is no surprise that almost all relativists fall in love with her. I am old enough not to do so. Besides divorces are quite common today. So we are in accordance with the spirit of the age when we leave geometry single as a convention. The gravitational field on the other hand is represented by the Christoffel symbols satisfying Einstein's equations. The Christoffel symbols come from a metric, but the latter has no geometric meaning.
In previous papers [1] [2] we have studied static spherically symmetric solutions of Einstein's equations on this ground. In contrast to the standard theory we have found solutions with arbitrary circular velocity dependence $V(r)$ which are of interest in connection with the dark matter problem.

Since most galaxies are not spherically symmetric we now investigate the axial symmetric situation. The corresponding standard vacuum solution is the Kerr metric. We use Boyer-Lindquist coordinates $(t,r,\te,\phi)$ which are identical with the spherical coordinates in the Schwarzschild case and have a clear operational physical definition [1]. We assume the line element in the form of Chandrasekhar [3], section 52
$$ds^2=e^{2\nu}dt^2-e^{2\psi}(d\phi -\omega dt)^2-e^{2\mu_2}dr^2-e^{2\mu_3}d\te^2\eqno(1.1)$$
with five metric functions $\nu,\psi,\omega,\mu_2,\mu_3$ which depend on $r,\te$, only. In the nonstandard theory it is important to choose the metric as general as possible. For this reason we give preference to the ansatz (1.1) compared with other ones in the literature which contain less than five functions. 

Chandrasekhar gives the Riemann tensor for (1.1) but he does not give the Ricci tensor $R_{\mu\nu}$. Instead he gives suitable combinations of the vacuum field equations $R_{\mu\nu}=0$, which allow integration of the resulting differential equations. We fully agree with all results of Chandrasekhar which finally lead to the Kerr metric. However, in his combination of the field equations some information gets lost; in particular the second derivatives of the functions $\mu_2$ and $\mu_3$ are eliminated. Then it must be checked whether the Kerr metric really satisfies {\it all} field equations. Such a check is done in sect.4. Our results for the Ricci tensor in sect.3 are also needed for future studies of the problem with matter. In sect.5 we construct a class of vacuum solutions with arbitrary angular velocity dependence $\Omega(r)$ in the equatorial plane. These solutions can be used to describe the outer part of spiral galaxies where normal matter can be neglected.

\section{Christoffel symbols}

Chandrasekhar has used differential forms to calculate the Riemann tensor. We find the route via Christoffel symbols most simple, and we need them anyway for the geodesic equations. According to (1.1) the non-vanishing components of the metric tensor are
$$ g_{tt}=e^{2\nu}-\omega^2 e^{2\psi},\quad g_{\phi\phi}=-e^{2\psi},\quad g_{t\phi}=\omega e^{2\psi}$$
$$g_{rr}=-e^{2\mu_2},\quad g_{\te\te}=-e^{2\mu_3}.\eqno(2.1)$$
Then the inverse metric is equal to
$$g^{tt}=e^{-2\nu},\quad g^{\phi\phi}=-e^{-2\psi}+w^2e^{-2\nu},\quad e^{t\phi}=\omega e^{-2\nu}$$
$$g^{rr}=-e^{-2\mu_2},\quad g^{\te\te}=-e^{-2\mu_3}.\eqno(2.2)$$

The Christoffel symbols are given by
$$\Gamma^\al_{\beta\gamma}={1\over 2}g^{\al\mu}(g_{\beta\mu},_\gamma
+g_{\gamma\mu},_\beta-g_{\beta\gamma},_\mu),\eqno(2.3)$$
the commas always mean partial derivatives. The non-vanishing components with upper index $t$ are
$$\Gamma^t_{rt}=\nu,_r-\omega,_r{\omega\over 2}e^{2(\psi-\nu)}$$
$$\Gamma^t_{\te t}=\nu,_\te-\omega,_\te{\omega\over 2}e^{2(\psi-\nu)}$$
$$\Gamma^t_{r\phi}=\eh e^{2(\psi-\nu)}\omega,_r\eqno(2.3)$$
$$\Gamma^t_{\te\phi}=\eh e^{2(\psi-\nu)}\omega,_\te.$$
With upper index $r$ we have
$$\Gamma^r_{tt}=\eh e^{-2\mu_2}(e^{2\nu}-\omega^2e^{2\psi}),_r$$
$$\Gamma^r_{\phi t}=\eh e^{-2\mu_2}(\omega e^{2\psi}),_r$$
$$\Gamma^r_{rr}=\mu_2,_r,\quad \Gamma^r_{r\te}=\mu_2,_\te$$
$$\Gamma^r_{\te\te}=-\eh e^{-2\mu_2}(e^{2\mu_3}),_r\eqno(2.4)$$
$$\Gamma^r_{\phi\phi}=-\eh e^{-2\mu_2}(e^{2\psi}),_r.$$
With upper index $\te$ we similarly get
$$\Gamma^\te_{tt}=\eh e^{-2\mu_3}(e^{2\nu}-\omega^2e^{2\psi}),_\te$$
$$\Gamma^\te_{\phi t}=\eh e^{-2\mu_3}(\omega e^{2\psi}),_\te$$
$$\Gamma^\te_{r\te}=\mu_3,_r,\quad \Gamma^\te_{\te\te}=\mu_3,_\te$$
$$\Gamma^\te_{rr}=-\eh e^{-2\mu_3}(e^{2\mu_2}),_\te\eqno(2.5)$$
$$\Gamma^\te_{\phi\phi}=-\eh e^{-2\mu_3}(e^{2\psi}),_\te.$$
Finally with upper index $\phi$ we have 
$$\Gamma^\phi_{rt}=\omega(\nu-\psi),_r-{\omega,_r\over 2}(1+\omega^2e^{2(\psi-\nu)})$$
$$\Gamma^\phi_{\te t}=\omega(\nu-\psi),_\te-{\omega,_\te\over 2}(1+\omega^2e^{2(\psi-\nu)})$$
$$\Gamma^\phi_{\phi r}=\psi,_r+{\omega\over 2}\omega,_re^{2(\psi-\nu)}\eqno(2.6)$$
$$\Gamma^\phi_{\phi\te}=\psi,_\te+{\omega\over 2}\omega,_\te e^{2(\psi-\nu)}.$$
As a consequence of axial symmetry we note the symmetry under simultaneous exchange of $r\leftrightarrow\te$
and $\mu_2\leftrightarrow\mu_3$.

\section{Ricci tensor}

Next we calculate the Ricci tensor according to
$$R_{\mu\nu}=\d_\al\Gamma^\al_{\mu\nu}-\d_\nu\Gamma^\al_{\mu\al}+\Gamma^\al_{\al\beta}\Gamma^\beta_{\mu\nu}
-\Gamma^\al_{\nu\beta}\Gamma^\beta_{\mu\al}.\eqno(3.1) $$
We perform this calculation in a suitable order to reconstruct the elimination of second derivatives done by Chandrasekhar.
$$R_{r\te}=\d_r\Gamma^r_{r\te}-\d_\te(\Gamma^t_{rt}+\Gamma^r_{rr})+\Gamma^r_{r\te}(\Gamma^t_{tr}+\Gamma^r_{rr}
+\Gamma^\phi_{r\phi})$$
$$+\Gamma^\te_{r\te}(\Gamma^t_{t\te}+\Gamma^r_{r\te}+\Gamma^\phi_{\phi\te})-\Gamma^t_{t\te}\Gamma^t_{rt}-\Gamma^t_{\te\phi}
\Gamma^\phi_{rt}-\Gamma^r_{r\te}\Gamma^r_{rr}-\Gamma^r_{\te\te}\Gamma^\te_{rr}-\Gamma^\phi_{t\te}\Gamma^t_{r\phi}
-\Gamma^\phi_{\te\phi}\Gamma^\phi_{r\phi}$$
$$=-(\psi+\nu),_r,_\te+\mu_2,_\te(\psi+\nu),_r+\mu_3,_\te(\psi+\nu),_\te-$$
$$-\psi,_r\psi,_\te-\nu,_r\nu,_\te+{1\over 2}\omega,_r\omega,_\te e^{2(\psi-\nu)}.
\eqno(3.2)$$
This agrees with equation (8) (p.274) of Chandrasekhar.

$$R_{\phi\phi}=\d_r\Gamma^r_{\phi\phi}+\d_\te\Gamma^\te_{\phi\phi}+\Gamma^r_{\phi\phi}(\Gamma^t_{tr}+\Gamma^r_{rr}
+\Gamma^\te_{r\te})$$
$$+\Gamma^\te_{\phi\phi}(\Gamma^t_{t\te}+\Gamma^r_{r\te}+\Gamma^\te_{\te\te})-2\Gamma^t_{r\phi}\Gamma^r_{t\phi}
-2\Gamma^\te_{t\phi}\Gamma^t_{\phi\te}$$
$$=-e^{2(\psi-\mu_2)}[\psi,_r,_r+\psi,_r(\psi+\nu-\mu_2+\mu_3),_r]-{1\over 2}\omega,_r^2e^{2(2\psi-\nu-\mu_2)}$$
$$+[r\leftrightarrow\te, \mu_2\leftrightarrow\mu_3].\eqno(3.3)$$
This agrees with equation (6) of Chandrasekhar. The last square bracket denotes the corresponding $\te$-terms which are present because of axial symmetry. If we set $R_{\phi\phi}=0$ we get
$$\psi,_r,_r+\psi,_\te,_\te=-\psi,_r(\psi+\nu-\mu_2+\mu_3),_r-{\omega,_r^2\over 2}e^{2(\psi-\nu)}$$
$$+[r\leftrightarrow\te, \mu_2\leftrightarrow\mu_3]\eqno(3.4)$$

$$R_{t\phi}=\d_r\Gamma^r_{t\phi}+\d_\te\Gamma^\te_{t\phi}+\Gamma^r_{t\phi}(\Gamma^r_{rr}+\Gamma^\te_{\te r})
+\Gamma^\te_{t\Phi}(\Gamma^r_{r\te}+\Gamma^\te_{\te\te})$$
$$-\Gamma^r_{tt}\Gamma^t_{r\phi}-\Gamma^\te_{tt}\Gamma^t_{\phi\te}-\Gamma^\phi_{tr}\Gamma^r_{\phi\phi}
-\Gamma^\Phi_{t\te}\Gamma^\te_{\phi\phi}.$$
$$={1\over 2}e^{2(\psi-\mu_2)}[\omega,_r,_r+2\omega\psi,_r,_r+\omega,_r(3\psi-\nu-\mu_2+\mu_3),_r+$$
$$+2\omega\psi,_r(\psi+\nu-\mu_2+\mu_3),_r]+{\omega\over 2}\omega,_r^2e^{2(2\psi-\nu-\mu_3)}$$
$$+[r\leftrightarrow\te, \mu_2\leftrightarrow\mu_3].\eqno(3.5)$$
Substituting (3.4) inhere we obtain equation (7) of Chandrasekhar:
$$e^{-2\mu_2}\omega,_r,_r+e^{-2\mu_3}\omega,_\te,_\te=-e^{-2\mu_2}\omega,_r(3\psi-\nu-\mu_2+\mu_3),_r$$
$$+[r\leftrightarrow\te, \mu_2\leftrightarrow\mu_3]\eqno(3.6)$$

Next we calculate
$$R_{tt}=\d_r\Gamma^r_{tt}+\d_\te\Gamma^\te_{tt}+\Gamma^r_{tt}(\Gamma^r_{rr}+\Gamma^\te_{r\te}+\Gamma^\phi_{r\phi})$$
$$+\Gamma^\te_{tt}(\Gamma^r_{r\te}+\Gamma^\te_{\te\te}+\Gamma^\phi_{\phi\te})-\Gamma^t_{tr}\Gamma^r_{tt}
-2\Gamma^r_{t\phi}\Gamma^\phi_{tr}-\Gamma^\te_{tt}\Gamma^t_{t\te}-2\Gamma^\te_{t\phi}\Gamma^\phi_{t\te}$$
$$=e^{-2\mu_2}\B[(\nu,_r,_r+2\nu,_r^2)e^{2\nu}-e^{2\psi}(\omega\omega,_r,_r+\omega,_r^2+4\omega\omega,_r\psi,_r
+\omega^2\psi,_r,_r+2\omega^2\psi,_r^2)\B]$$
$$+e^{2(\psi-\mu_2)}(\omega,_r+2\omega\psi,_r)\B[{\omega,_r\over 2}(1+\omega^2e^{2(\psi-\nu)})+\omega(\psi-\nu),_r\B]$$
$$+e^{-2\mu_2}\B[\nu,_re^{2\nu}-\omega e^{2\psi}(\omega,_r+\omega\psi,_r)\B]\B[(\psi-\nu+\mu_3-\mu_2),_r+\omega\omega,_r
e^{2(\psi-\nu)}\B]$$
$$+[r\leftrightarrow\te, \mu_2\leftrightarrow\mu_3]\eqno(3.7)$$
Using (3.4) and (3.6) we can eliminate the second derivatives of $\omega$ and $\psi$. Then Einstein's equation $R_{tt}=0$ implies
$$0=e^{-2\mu_2}\B[\nu,_r,_r+\nu,_r(\psi+\nu+\mu_3-\mu_2),_r\B]-{\omega,_r^2\over 2}e^{2(\psi-\nu-\mu_2)}$$
$$+[r\leftrightarrow\te, \mu_2\leftrightarrow\mu_3].\eqno(3.8)$$
This is equation (5) of Chandrasekhar.

The remaining non-vanishing components are $R_{rr}$ and $R_{\te\te}$ and these are most subtle because they contain second derivatives of four functions. We have
$$R_{rr}=\d_\te\Gamma^\te_{rr}-\d_r(\Gamma^t_{rt}+\Gamma^\te_{r\te}+\Gamma^\phi_{r\phi})+\Gamma^r_{rr}(\Gamma^t_{tr}+\Gamma^\te_{r\te}
+\Gamma^\phi_{r\phi})$$
$$+\Gamma^\te_{rr}(\Gamma^t_{t\te}+\Gamma^\te_{\te\te}+\Gamma^\phi_{\phi\te})-(\Gamma^t_{rt})^2-2\Gamma^t_{r\phi}\Gamma^\phi_{tr}$$
$$-\Gamma^r_{r\te}\Gamma^\te_{rr}-(\Gamma^\te_{r\te})^2-(\Gamma^\phi_{r\phi})^2$$
$$=-e^{2(\mu_2-\mu_3)}\B[\mu_2,_\te,_\te+2\mu_2,_\te(\mu_2-\mu_3),_\te\B]-\mu_3,_r,_r-\d^2_r(\psi+\nu)$$
$$-\mu^2_3,_r-\psi,_r^2-\nu,_r^2+\mu_2,_r(\psi+\nu+\mu_3),_r-$$
$$-e^{2(\mu_2-\mu_3)}\mu_2,_\te(\psi+\nu+\mu_3-\mu_2),_\te+{1\over 2}\omega,_r^2e^{2(\psi-\nu)}.\eqno(3.9)$$
$R_{\te\te}$ is obtained from $R_{rr}$ by the substitutions $r\leftrightarrow\te$ $\mu_2\leftrightarrow\mu_3$ as before.
Multiplying $R_{\te\te}$ by $\exp[2(\mu_2-\mu_3)]$ and subtracting it from (3.9) the second derivatives of $\mu_2$ and $\mu_3$ drop out. Using Einstein's equations we arrive at the following simpler equation
$$e^{-2\mu_3}\B[\d^2_\te(\psi+\nu)+\psi,_\te^2+\nu,_\te^2-(\psi+\nu),_\te(\mu_2+\mu_3),_\te\B]-{\omega,_\te^2\over 2} e^{2(\psi-\nu-\mu_3)}=$$
$$=e^{-2\mu_2}\B[\d^2_r(\psi+\nu)+\psi,_r^2+\nu,_r^2-(\psi+\nu),_r(\mu_3+\mu_2),_\te\B]+{\omega,_r^2\over 2}e^{2(\psi-\nu-\mu_2)}.
\eqno(3.10)$$
On the other hand adding (3.3) and (3.8) we get
$$e^{-2\mu_3}\d^2_\te(\psi+\nu)+e^{-2\mu_3}(\psi+\nu),_\te(\psi+\nu+\mu_2-\mu_3),_\te=$$
$$=-e^{-2\mu_2}\d^2_r(\psi+\nu)-e^{-2\mu_2}(\psi+\nu),_r(\psi+\nu+\mu_3-\mu_2),_r.\eqno(3.11).$$
Combining this with (3.10) we find Chandrasekhar's last two equations (9) and (10). But it is clear from the derivation of (3.10) that the six differential equations of Chandrasekhar contain less information than the six Einstein's equations. Therefore we must check at the end that all Einstein's equations are satisfied, in particular $R_{rr}=0$.

\section{Calculation of $R_{rr}$}

The six differential equations of Chandrasekhar can be integrated and finally lead to the Kerr metric [1]. Let $M$ and $a$ be mass and angular momentum of the Kerr black hole, then the five metric functions in (1.1) are given by
$$\ro^2=r^2+a^2\cos^2\te,\quad \Delta=r^2-2Mr+a^2\eqno(4.1)$$
$$\Sigma^2=(r^2+a^2)^2-a^2\Delta\sin^2\te\eqno(4.2)$$
$$\omega={2aMr\over\Sigma^2}\eqno(4.3)$$
$$e^{2\psi}={\Sigma^2\over\ro^2}\sin^2\te\eqno(4.4)$$
$$e^{2\nu}={\ro^2\Delta\over\Sigma^2}\eqno(4.5)$$
$$e^{2\mu_2}={\ro^2\over\Delta},\quad e^{2\mu_3}=\ro^2.\eqno(4.6)$$
It is straightforward but quite cumbersome to calculate $R_{rr}$ (3.9) with these results. The first derivatives of the 5 functions have been given by Chandrasekhar on p.290. We need the following second derivatives
$$\d^2_r(\psi+\nu)={1\over\Delta}-2{(r-M)^2\over\Delta^2}$$
$$\d^2_\te\mu_2=-{a^2\over\ro^2}\cos 2\te-{a^4\over 2\ro^4}\sin^2 2\te\eqno(4.7)$$
$$\d^2_r\mu_3={1\over\ro^4}(a^2\cos^2\te-r^2).$$

Substituting everything into (3.9) we obtain
$$e^{2\mu_3}R_{rr}={1\over\Delta}\B({a^4\over 2\ro^2}\sin^2 2\te+a^2\cos 2\te\B)-\ro^2\B[{1\over\ro^4}(a^2\cos^2\te-r^2)
+{1\over\Delta}-2{(r-M)^2\over\Delta^2}+$$
$$+{r^2\over\ro^4}+\nu^2,_r+\psi^2,_r-\B({r\over\ro^2}-{r-M\over\Delta}\B)\B({r-M\over\Delta}+{r\over\ro^2}\B)\B]+$$
$$+{a^2\over 2\Delta}\sin 2\te\cot\te+{2a^2M^2\over\Sigma^8}[(r^2+a^2)(3r^2-a^2)-a^2(r^2-a^2)\sin^2\te]^2\times$$
$$\times\sin^2\te\B(r^2+a^2+2a^2Mr{\sin^2\te\over\ro^2}\B){\Sigma^2\over\Delta}.\eqno(4.8)$$
We use
$$\psi^2,_r+\nu^2,_r=(\psi,_r+\nu,_r)^2-2\psi,_r\nu,_r$$
and
$$2\ro^2\psi,_r\nu,_r=2\B\{{\ro^2\over\Sigma^2}[2r(r^2+a^2)-a^2(r-M)\sin^2\te]-r\B\}\times$$
$$\times\B\{-{1\over\Sigma^2}[2r(r^2+a^2)-a^2(r-M)\sin^2\te]+{r\over\ro^2}+{r-M\over\Delta}\B\}.$$
Then we collect all terms on the common denominator $\Sigma^6\Delta^2\ro^2$:
$$e^{2\mu_3}R_{rr}\Sigma^6\Delta^2\ro^2=a^2\Delta\Sigma^6(\eh a^2\sin^2 2\te+\ro^2\cos 2\te)+\Sigma^6\Delta^2(r^2-a^2\cos^2\te)+$$
$$+\ro^4\Sigma^6(r^2-2Mr+2M^2-a^2)-2\Sigma^6\ro^4(r-M)^2+$$
$$+2\{\Sigma^2\Delta\ro^2[2r(r^2+a^2)-a^2(r-M)\sin^2\te]-r\Sigma^4\Delta\}\times$$
$$\times\{-\Delta\ro^2[2r(r^2+a^2)-a^2(r-M)\sin^2\te]+r\Sigma^2\Delta+(r-M)\Sigma^2\ro^2\}+$$
$$+a^2\Sigma^6\Delta\ro^2\cos^2\te
+2a^2M^2\Delta\sin^2\te[(r^2+a^2)(3r^2-a^2)-a^2(r^2-a^2)\sin^2\te]^2\times$$
$$\times[\ro^2(r^2+a^2)+2a^2Mr\sin^2\te].\eqno(4.9)$$

To minimize the probability of a computational error we have computed the polynomial on the r.h.s. of (4.9) by means of the algebraic computer program FORM [4]. There is a huge cancellation and the r.h.s. is indeed zero. Then by symmetry $R_{\te\te}$ vanishes as well. Hence the Kerr metric satisfies all Einstein's equations. This closes a loop-hole in the treatment by Chandrasekhar. 

\section{The Schwarzschild and Kerr classes}

Chandrasekhar in his solution of the differential equations makes some simplifications of the integration procedure by choosing a suitable gauge. In the nonstandard theory we want a physical gauge fixing, for example by giving the rotation curve $V(r)$ in the equatorial plane. To integrate the field equations with such a constraint is a formidable task. Fortunately this can be completely avoided by using gauge (or diffeomorphism) invariance. In [1] we have found in the Schwarschild case that the solutions found by integrating the field equations can also be obtained by a suitable gauge transformation from the Schwarzschild solution. For better understanding of the following we repeat the argument here. We start from the Schwarzschild metric
$$ds^2=\B(1-{r_s\over\bar r}\B)d\bar t^2-\B(1-{r_s\over\bar r}\B)^{-1}d\bar r^2-\bar r^2(d\te^2+\sin^2\te d\phi^2)
\eqno(5.1)$$ 
with a fictitious unphysical radial coordinate $\bar r$. We set our physical, given circular velocity squared $u(r)=V^2(r)$ equal to the Schwarzschild expression
$$u(r)={r_s\over 2(\bar r-r_s)}.\eqno(5.2)$$
where $r_s=2M$ is the Schwarschild radius. Then solving for $\bar r$ gives the desired diffeomorphism
$$\bar r=r_s{2u(r)+1\over 2u(r)}.\eqno(5.3)$$
Indeed we get
$$g_{tt}=1-{r_s\over\bar r}={1\over 2u+1}$$
$$g_{\te\te}=\bar r^2=r_2^2\B({2u+1\over 2u}\B)^2$$
$$g_{rr}=\B(1-{r_s\over\bar r}\B)^{-1}\B({d\bar r\over dr}\B)^2=\B({r_s\over 2}\B)^2(2u+1)\B({u'\over u^2}\B)^2\eqno(5.4)$$
which is the class of nonstandard solutions found in [1].

Instead of using the circular velocity we can work with the angular velocity
$$\bar\Omega(\bar r)={d\phi\over dt}={\sqrt{M}\over \bar r^{3/2}}.\eqno(5.5)$$ 
Then the corresponding diffeomorphism is
$$\bar r=\B({M\over\Omega^2}\B)^{1/3}\eqno(5.6)$$
and the Schwarzschild class gets the following form
$$g_{tt}=1-(8M^2\Omega^2)^{1/3}$$
$$g_{\te\te}=\B({M\over\Omega^2}\B)^{2/3}$$
$$g_{rr}={4\over 9}[1-(8M^2\Omega^2)^{1/3}]^{-1}\B({M\over\Omega^5}\B)^{2/3}(\Omega')^2.\eqno(5.6)$$
Here $\Omega(r)$ is the prescribed angular velocity. 

In the Kerr case the angular velocity in the equatorial plane is equal to ([3], p.336)
$$\bar\Omega={\pm\sqrt{M}\over\bar r^{3/2}\pm a\sqrt{M}}.\eqno(5.7)$$
This depends on $\bar r$ only, so that with this observable the Kerr class becomes as simple as the Schwarzschild class. Solving for $\bar r$ gives the required diffeomorphism
$$\bar r=M^{1/3}\B({1\over\Omega(r)}-a\B)^{2/3}\eqno(5.8)$$
and 
$${d\bar r\over dr}=-{2\over 3}M^{1/3}\B({1\over\Omega}-a\B)^{-1/3}{\Omega'\over\Omega^2}.\eqno(5.9)$$
Then the quantities (4.1) (4.2) giving the Kerr metric must be changed according to
$$\bar\ro^2=M^{2/3}\B({1\over\Omega}-a\B)^{4/3}+a^2\cos^2\te\eqno(5.10)$$
$$\bar\Delta=M^{2/3}\B({1\over\Omega}-a\B)^{4/3}-2M^{4/3}\B({1\over\Omega}-a\B)^{2/3}+a^2\eqno(5.11)$$
$$\bar\Sigma^2=\B[M^{2/3}\B({1\over\Omega}-a\B)^{4/3}+a^2\B]^2-a^2\bar\Delta\sin^2\te.\eqno(5.12)$$
The quantities (4.3-6) must now be calculated with these new expressions with bar and with $\bar r$. Finally the modified axial symmetric metrics follow from (2.1) where $g_{rr}$ gets the factor (5.9) squared. This gives the Kerr class. These metrics are physically different from the Kerr metric and have a better chance to describe the outer part of spiral galaxies where matter can be neglected.

\end{document}